\pdfoutput=1
\documentclass[12pt]{article}
\usepackage[utf8]{inputenc} 
\usepackage[T1]{fontenc}
\usepackage{hyperref} 
\usepackage{orcidlink} 
\usepackage{url}      
\usepackage{setspace}
\usepackage{dcolumn}
\usepackage[margin=1in]{geometry}
\usepackage{amsmath}
\DeclareMathOperator{\sech}{sech} 
\usepackage{graphicx}
\usepackage{subcaption} 
\usepackage{float}
\usepackage{tabularx}
\usepackage[style=numeric-comp, sorting=none]{biblatex} 
\bibliography{reference.bib}
\usepackage[mathscr]{eucal}
\usepackage{multirow}
\usepackage{booktabs}
\usepackage[dvipsnames]{xcolor} 
\usepackage{chngcntr} 
\counterwithin{equation}{section} 

\usepackage{xcolor}

\DeclareFieldFormat[article]{title}{\textcolor{magenta}{#1}}

\AtEveryBibitem{%
  \ifentrytype{article}{%
    \renewbibmacro*{journal+issuetitle}{%
      {\color{blue}
        \printfield{journaltitle}%
        \setunit*{\addspace}%
        \printfield{volume}%
        \printfield{number}%
        \setunit{\addspace}%
        \printfield{year}}%
    }%
    \renewbibmacro*{pagination}{%
      \setunit{\addcomma\space}%
      {\color{blue}\printfield{pages}}%
    }%
  }{}%
}

\begin{document}

\title{\textbf{\textcolor{BrickRed}{ {\Large Inflationary models in a minimally coupled $f(R,T)$ gravity: Constraints from $Planck$, BICEP/$Keck$, and ACT}
}}}

\author{ \textcolor{Violet}{\textbf{ Biswajit Deb}} \orcidlink{0000-0001-8992-7600} \footnote{Electronic Address: \textcolor{blue}{biswajit.deb@aus.ac.in}} , \textcolor{Violet}{\textbf{Atri Deshamukhya}} \orcidlink{0000-0003-4350-6645} \footnote{Electronic Address: \textcolor{blue}{atri.deshamukhya@aus.ac.in}}\\
Department of Physics, Assam University, Silchar, India }

\date{}
\maketitle

\begin{abstract}
The advent of high-precision cosmological observations has challenged many traditional inflationary models. Data from $Planck$ 2018 along with the BICEP/$Keck$ 2018 result have already ruled out most of the established models by placing tight constraints on the tensor-to-scalar ratio $r$. Upcoming missions like LiteBIRD \& CMB-S4 are expected to impose an even more stringent bound on $r$ , potentially excluding further models from the viable landscape. In this evolving observational context, modified gravity theories offer a promising way to reconcile inflationary models with data. In this work, we explore several inflationary models, namely mutated hilltop inflation, D-brane inflation, and Woods-Saxon inflation, within the framework of $f(R,T)$ gravity. A minimally coupled and linear combination of Ricci scalar and trace of EM tensor is considered as $f(R,T)=R+16\pi G \lambda T$ and the cosmological observable parameters, viz. scalar spectral tilt $n_s$, tensor-to-scalar ratio $r$, and running of scalar spectral index $n_{sk}$ are estimated for the three models, and their trajectories are plotted in the $n_s-r$ plane. The model results are evaluated in light of the $Planck$, BICEP/$Keck$, DESI DR2, and ACT DR6 data. We observe that for a certain model parameter space, these potentials are viable within the current observational bounds. \\


\end{abstract}

\section{Introduction}	
Theoretical cosmology went through a paradigm shift after Alan Guth proposed cosmic inflation theory to describe the post-Planck Universe \cite{R1}. Inflation not only addressed several flaws of the standard Big-Bang model, viz. causality and flatness problems, but also provided a natural mechanism for generating the quantum fluctuations that seed the formation of large-scale structures of the Universe \cite{R2, R3, R4}. In addition, inflation bridges cosmology with particle physics, thereby allowing a smooth transition from radiation to the matter-dominated Universe \cite{R5, R6, R7}. These unique features of inflation establish it as a fundamental pillar of the standard model of cosmology.  Over time, the primordial fluctuations generated during inflation develop into the anisotropies detected in the Cosmic Microwave Background (CMB), and are quantitatively characterized by key observable parameters such as the scalar spectral index $(n_s)$, tensor-to-scalar ratio $(r)$, and running of the scalar spectral index $(n_{sk})$ \cite{R5, R6, R7}. \\ \\
The de Sitter expansion during inflation is usually governed by a scalar field, called the inflaton $\phi$, which rolls over its potential as inflation proceeds \cite{R5,R6,R7}. Over the past decades, a plethora of potentials motivated from phenomenology, string theory, supper symmetry, quantum field theory, etc. have been used to develop inflationary models that can match observational data from the CMB probes \cite{R8}. \\ \\
The advanced and precise measurement of CMB anisotropies by $Planck$ left most theoretical models in doubt \cite{R9,R10}. The latest data release from the Atacama Cosmology Telescope (ACT) brought about even more turbulence in the validity of inflationary models \cite{R11,R12}. ACT DR6 provides a larger value of $n_s$, ruling out a broad class of inflationary models at the $2\sigma$ confidence level, which includes Starobinsky inflation, Higgs inflation, and $\alpha$-attractors. Whereas power-law inflation is supported by ACT data at $1\sigma$ confidence level, which was earlier ruled out by $Planck$ \cite{R9, R11}. Recent works can be checked in Ref. \cite{R13,R14,R15,R16,R17,R18,R19,R20,R21,R22,R23,R24,R25,R26,R27,R28,R29,R30} for more insight into this matter. \\ \\
The scalar spectral tilt reported by $Planck$ 2018 \cite{R9} is $n_s = 0.9651 \pm 0.0044$ while reported by ACT \cite{R11}, it is $n_s=0.9666 \pm 0.0077$. However, a combined analysis of the $Planck$ and ACT data provides a larger scalar spectral tilt
$n_s = 0.9709 \pm 0.0038$ (P-ACT) \cite{R11}. When CMB lensing and Baryon Acoustic Oscillation (BAO) distance measurements from the Dark Energy Spectroscopic Instrument (DESI)\cite{R31, R32} are included, the value increases to $n_s = 0.9743 \pm 0.0034$ (P-ACT-LB), which deviates from the original Planck result by $2\sigma$ \cite{R11}. Then by adding B-mode polarization data from the BICEP/Keck telescopes (BK18) \cite{R33} to this combined data set, the upper limit of the tensor-to-scalar ratio slightly elevates to $r < 0.038$ (P-ACT-LB-BK18) at the $95\%$ confidence level \cite{R12}. This tension between different CMB probe results demands the inflationary models to be revisited.   \\ \\
General relativity (GR) cannot accommodate all evolutionary phases simultaneously and fails to explain the dark sector despite its tremendous success at low-energy scales and in the weak-field limit \cite{R34}. These flaws motivated theorists to look beyond GR and to make corrections to it \cite{R35}. The $f(R)$ theory or the $R^2$ inflation is the first of its kind proposed by Starobinsky \cite{R36}. Subsequently several modified gravity theories came up and yield excellent results in case of early to late-time era, dark matter, dark energy, and compact objects etc. Inflation has also been studied in different modified gravity theories, viz. $f(R)$, $f(R,T)$, $f(G)$, $f(R,G)$, $f(\phi,T)$, etc. where $R, T, G$ stands for the Ricci scalar, trace of the EM tensor, and Gauss-Bonnet scalar \cite{R36,R37,R38,R39,R40,R41,R42,R43,R44,R45,R46,R47}. \\ \\
Harko et al. proposed ${f(R,T)}$  gravity, in which the trace of the EM tensor was included in the Lagrangian \cite{R48}. $f(R,T)$ gravity not only satisfies the necessary energy conditions \cite{R49,R50}, it is quite sound in explaining different topics of cosmology ranging from dark energy, dark matter, black hole to wormhole, etc \cite{R51,R52,R53,R54,R55}. Inflation has been studied in $f(R,T)$ gravity with both minimal and non-minimal coupling cases \cite{R37, R38,R39,R40,R56}. Inflationary models like Power-Law inflation, Chaotic inflation, Natural inflation, etc. which were ruled out by $Planck$ in GR, were found to be consistent in $f(R,T)$ gravity \cite{R38,R41,R56,R57}. Warm inflation is also studied in $f(R,T)$ gravity that produced $r \sim 10^{-3}$ compatible with future CMB probes such as LiteBIRD \cite{R57}. All of these success stories make $f(R,T)$ gravity a prolific candidate for further exploration. \\ \\
In this work, we have taken three highly motivated inflationary models, viz. Mutated hilltop inflation, D-brane inflation, and Woods-Saxon inflation to study in $f(R,T)$ gravity. The effect of the correction term on inflationary observables like $n_s$, $r$, and $n_{sk}$ will be evaluated for three potentials. Furthermore, the predicted model results will be confronted with recent bounds from $Planck$, BICEP/$Keck$, ACT DR6, and DESI DR2 to check their viability. 
 \\ \\ 
The paper has been organized as follows: In Section \ref{s2}, inflation in ${f(R,T)}$ gravity will be reviewed briefly, and in Section \ref{s3}, we will present the perturbation calculations in case of $f(R,T)$ gravity. In Section \ref{s4}, we will discuss three models, viz. Mutated hilltop inflation, D-brane inflation, and Woods-Saxon inflation in detail with their results in $f(R,T)$ gravity. In Section \ref{s4c}, we checked the compatibility of the models with the latest CMB probe data from ACT DR6 and DESI DR2. In Section \ref{s5}, we present our conclusion. Throughout the manuscript, the natural unit system $c= \hbar = \kappa=1$, and the sign convention $(-,+,+,+)$  for the metric tensor are used. Where $\kappa=8\pi G=M_{Pl}^{-2}$, $M_{Pl}$ being the reduced $Planck$ mass.


\section{Inflation in {\it{f(R,T)}} gravity} \label{s2}
The ${f(R,T)}$ gravity action proposed by Harko et al. \cite{R48} reads as:
\begin{equation}
    S = \int \left[\frac{f(R,T)}{16 \pi G} + L_m  \right] \sqrt{-g} d^4x 
    \label{e1}
\end{equation}
 where ${f(R,T)}$ is an arbitrary function of the Ricci scalar $R$ and the trace of the energy-momentum tensor $T_{\alpha\beta}$, $L_m$ is the matter Lagrangian density, $G$ is the Newtonian gravitational constant, and $g$ is the metric determinant. On metric variation of the action, the ${f(R,T)}$ gravity field equations are obtained as
 \begin{equation}
    f_R (R,T) R_{\alpha\beta} - \frac{1}{2} g_{\alpha\beta}f(R,T) + [g_{\alpha\beta} \nabla_\sigma \nabla^\sigma - \nabla_\alpha \nabla_\beta] f_R (R,T) =  8\pi G T_{\alpha\beta} - f_T (R,T) (T_{\alpha\beta} + \Theta_{\alpha\beta})
    \label{e2}
\end{equation}
where we have denoted $f_R (R,T)= \frac{\partial f(R,T)}{\partial R}$ , $f_T (R,T)= \frac{\partial f(R,T)}{\partial T}$ and defined $T_{\alpha\beta}$ and $\Theta_{\alpha\beta}$ as
\begin{equation}
     T_{\alpha\beta} = g_{\alpha\beta}L_m - 2 \frac{\delta L_m}{\delta g^{\alpha\beta}}
     \label{e3}
\end{equation}
\begin{equation}
    \Theta_{\alpha\beta}= g^{\mu\nu} \frac{\delta T_{\mu\nu}}{\delta g^{\alpha\beta}} = -2 T_{\alpha\beta} + g_{\alpha\beta}L_m - 2 \frac{\delta^2 L_m}{\delta g^{\alpha\beta} \delta g^{\mu\nu}}
    \label{e4}
\end{equation}
The term $\Theta_{\alpha\beta}$ is crucial in ${f(R,T)}$ gravity, as it contains Lagrangian $L_m$. Depending on the nature of the matter field, the field equations for ${f(R,T)}$ gravity will vary. So, the field equations in ${f(R,T)}$ gravity depend not only on the choice of the function but also on the nature of the matter field \cite{R48}. \\ \\
In this work, we have considered the simplest yet most studied form of ${f(R,T)}$ which is $f(R,T) = R + 16 \pi G \lambda T$, where $\lambda$ is the model parameter. This particular functional form satisfies all energy conditions \cite{R49} and is excellent for explaining both inflation and late-time cosmology \cite{R38,R39,R40,R41,R58,R59}.  \\ \\
The field equations for this function will be
\begin{equation}
    R_{\alpha\beta} - \frac{1}{2}g_{\alpha\beta}R = 8 \pi G T_{\alpha\beta}^{(eff)}
\label{e5}
\end{equation}
where $T_{\alpha\beta}^{(eff)}$ is the effective EM tensor that includes the correction from $f(R,T)$ gravity, given by,
\begin{equation}
    T_{\alpha\beta}^{(eff)}= T_{\alpha\beta} - 2 \lambda( T_{\alpha\beta} - \frac{1}{2}T g _{\alpha\beta} + \Theta_{\alpha\beta})
\label{e6}
\end{equation}
For $\lambda=0$, the Einstein field equations are retrieved. To describe inflation, a homogeneous scalar field $\phi$ minimally coupled to gravity is required \cite{R5,R6,R7}. Then the matter Lagrangian will take the form,
\begin{equation}
    L_m = -\frac{1}{2}g^{\alpha\beta}\partial_{\alpha}\phi \partial_{\beta}\phi - V(\phi) = \frac{1}{2}\dot \phi^2 - V(\phi)
\label{e7}
\end{equation}
where $V(\phi)$ is the potential of the inflaton. To study the cosmological implications, we consider the Friedman-Lemaitre-Robertson-Walker (FLRW) metric for a flat Universe. Then from Eq. \ref{e5}, modified Friedmann equations are derived as:
\begin{equation}
    H^2 = \frac{8\pi G}{3}\left[\frac{\dot\phi^2}{2} (1+ 2\lambda) + V(\phi) (1+ 4\lambda)\right]
    \label{e8}
\end{equation}
\begin{equation}
    \frac{\ddot a}{a} = - \frac{8\pi G}{3} \left[\dot\phi^2 (1+ 2\lambda) - V(\phi) (1+ 4\lambda)\right]
    \label{e9}
\end{equation}
The Hubble parameter is defined as $H=\frac{\dot a}{a}$, and $a(t)$ is the scale factor. Likewise, the effective equation of state takes the form of
\begin{equation}
    w^{(eff)} = \frac{p^{(eff)}}{\rho^{(eff)}} = \frac{\dot\phi^2 (1+ 2\lambda) - 2V(\phi) (1+ 4\lambda)}{\dot\phi^2 (1+ 2\lambda) + 2V(\phi) (1+ 4\lambda)}
    \label{e10}
\end{equation}
Here, $p^{eff}$ and $\rho^{eff}$ are effective pressure and radiation density, respectively. However, Eq. \ref{e9} can be expressed in terms of $\dot H$ as:
\begin{equation}
    \dot H = \frac{\ddot a}{a} - H^2 = - \frac{8\pi G}{2} (p^{eff} + \rho^{eff})= - 4\pi G \dot\phi^2(1+ 2\lambda)
    \label{e11}
\end{equation}
Now, from Eqs. \ref{e8}, \ref{e9}, and \ref{e11} the modified Klein-Gordon equation is derived as:
\begin{equation}
    \ddot \phi (1+ 2\lambda) + 3H\dot \phi (1+ 2\lambda) + \frac{dV}{d\phi} (1+ 4\lambda) = 0 
    \label{e12}
\end{equation}
To check whether slow roll inflation is possible in this model, standard slow-roll parameters need to be derived for $f(R,T)$ gravity. The first slow-roll parameter denoted by $\epsilon$ in terms of the model parameter $\lambda$ is
\begin{equation}
    \epsilon_H= - \frac{\dot H}{H^2} = \frac{3 \dot\phi^2 (1+ 2\lambda)}{\dot\phi^2 (1+ 2\lambda) + 2 V(\phi) (1+ 4\lambda)}
    \label{e13}
\end{equation}
Accelerated expansion occurs in the slow roll regime if $\epsilon << 1$, that is, when the potential energy of the inflaton dominates the kinetic energy. Under this condition, the inflaton rolls slowly and hence the name slow-roll condition \cite{R5,R6,R7}. In terms of the current model parameter, the modified first slow roll condition reads as follows:
\begin{equation}
    \dot\phi^2 (1+ 2\lambda) << V(\phi) (1+ 4\lambda)
    \label{e14}
\end{equation}
Again, the accelerated expansion must sustain for a prolonged period of time to achieve the desired amount of $e$-folds, and hence a second condition must be imposed \cite{R5,R6,R7}. In terms of our model parameter this reads as follows:
\begin{equation}
    \ddot\phi (1+ 2\lambda) << | 3 H \dot \phi (1+ 2 \lambda)|
    \label{e15}
\end{equation}
This defines the second slow-roll condition $ |\eta| << 1 $ as:
\begin{equation}
    \eta_H= - \frac{\ddot\phi}{H \dot \phi}
    \label{e16}
\end{equation}
The parameters defined in Eqs. \ref{e13} and \ref{e16} are called Hubble slow-roll parameters. They can also be defined in terms of potential and are therefore called potential slow-roll parameters $\epsilon_{V}$ and $\eta_{V}$ \cite{R5}. The potential slow-roll parameters in $f(R,T)$ gravity in terms of reduced $Planck$ mass $M_{Pl}$ read:
\begin{align}
    \epsilon_V &= \frac{1}{1+ 2 \lambda} \frac{M_{Pl}^2}{2} \left(\frac{V_{\phi}}{V}\right)^2 \nonumber \\
    \eta_V &=  \frac{1}{1+ 2 \lambda} M_{Pl}^2 \frac{V_{\phi\phi}}{V}
    \label{e17}
\end{align}
Where $V_{\phi}=\frac{dV(\phi)}{d\phi}$ and $V_{\phi\phi}=\frac{d^2V(\phi)}{d\phi^2}$. It is clear from the Eqs. \ref{e17} that the affect of the correction term has been induced in the slow-roll parameters. As long as $\epsilon_V, |\eta_V| << 1$, slow-roll inflation continues and inflation ends as soon as either of the parameters becomes unity \cite{R7}. The third potential slow-roll parameter, which describes how fast the curvature of the potential is changing as the field evolves \cite{R5}, in $f(R,T)$ gravity reads:
\begin{equation}
    \xi_{V}^2= \frac{M_{Pl}^4}{(1+2\lambda)^2}\frac{V_{\phi} V_{\phi\phi\phi
    }}{V^2}
    \label{e18}
\end{equation}
Another important quantity required to study in any inflationary model is the amount of inflation that has taken place. It is quantified by the number of $e$-folds and is defined as
\begin{equation}
    N = \ln a = \int_{\phi_i}^{\phi_{end}} \frac{H}{\dot \phi} \,d\phi\
    \label{e19}
\end{equation}
where $\phi_{end}$ and $\phi_{i}$ indicate the value of the inflaton at the end and beginning of inflation \cite{R5}. Under slow-roll approximation, the number of $e$-folds in $f(R,T)$ gravity becomes
\begin{equation}
    N= \frac{1+2\lambda}{M_{Pl}^2} \int_{\phi_{end}}^{\phi_i} \frac{V(\phi)}{V_{\phi}(\phi)} \,d\phi\ 
    \label{e20}
\end{equation}
\section{Perturbations in $f(R,T)=R+16\pi G \lambda T $ gravity}\label{s3} 
During inflation, the perturbation in the metric tensor and inflaton field around the homogeneous background are expressed as 
\begin{equation}
    \begin{split}
        g_{\alpha\beta}(t,x) &= g_{\alpha\beta}(t)+\delta g_{\alpha\beta}(t,x)   \\
        \phi(t,x) & =\phi_0(t)+\delta\phi(t,x) 
    \label{e21a}
    \end{split}
\end{equation}
Here $\delta g_{\alpha \beta}(t,x)$ and $\delta\phi (t,x)$ represent the perturbations in the metic and field. The linear order perturbed FLRW metric for flat space in Longitudinal guage (Newtonian guage) reads \cite{R5}
\begin{equation}
    ds^2= -(1+2\Phi)dt^2+a^2(t)(1-2\Psi)\delta_{ij}dx^idx^j \label{21b}
\end{equation}
where $\Phi$ and $\Psi$ are the two independent functions that describe the scalar perturbations. In linear order perturbation, scalar fields do not possess anisotropic stresses, and hence $\Phi=\Psi$ \cite{R5, R59d}. Then the components of perturbed Einstein equations are as follows:
\begin{equation}
    3H(\dot\Phi+H\Phi)-\frac{\nabla^2\Phi}{a^2}=-4\pi G\delta\rho^{eff}
    \label{21c}
\end{equation}
\begin{equation}
    \dot\Phi+H\Phi=4 \pi G(1+2\lambda)\dot \phi_0 \delta\phi
    \label{21d}
\end{equation}
\begin{equation}
    \ddot\Phi+4H\dot\Phi+(2\dot H+3H^2)\Phi=4\pi G\delta p^{eff}
    \label{21e}
\end{equation}
where $\delta\rho^{eff}$ and $\delta p^{eff}$ denote the corrected perturbed energy density and pressure in $f(R,T)$ gravity given by
\begin{equation}
    \delta\rho^{eff}=(1+2\lambda)\delta\rho+2\lambda V_{\phi} \delta\phi
    \label{21f}
\end{equation}
\begin{equation}
    \delta p^{eff}=(1+2\lambda)\delta p-2\lambda V_{\phi} \delta\phi
    \label{21g}
\end{equation}
The differential equation for the Bardeen potential $\Phi$ can be derived using the perturbed Einstein Eqs. \ref{21c} and \ref{21e} as
\begin{equation}
    \Phi''+ 3\mathcal{H}(1+c_{a}^2)\Phi'+ \left[2\mathcal{H'}+(1+3c_{a}^{2})\mathcal{H}^2 \right]\Phi -c_{a}^2\nabla^2\Phi=4\pi Ga^2 \delta p_{nad}
    \label{21h}
\end{equation}
Where prime denotes differentiation with respect to the conformal time $\tau$ and $\mathcal{H}=\frac{a'}{a}=aH$ is the conformal Hubble parameter. Furthermore, $\delta p_{nad}$ is the gauge invariant non adiabatic pressure perturbation defined as \cite{R59a, R59d}
\begin{equation}
    \delta p_{nad}=\delta p^{eff} - c_{a}^2 \delta \rho^{eff}
    \label{21i}
\end{equation}
where $c_{a}^2=\delta p^{eff}/\delta \rho^{eff}$ denotes the adiabatic speed of sound in the case of $f(R,T)$ gravity. Again, in scalar filed models the non adiabatic pressure perturbation can be expressed as a difference between the effective speed of sound $c_e^2$ and the adiabatic speed of sound $c_a^2$ as \cite{R59a}:
\begin{equation}
    \delta p_{nad}= \left(\frac{c_e^2-c_a^2}{4\pi G a^2}\right) \nabla^2 \Phi 
    \label{21j}
\end{equation}
The effective speed of sound is defined by Garriga and Mukhanov \cite{R59b} for the scalar field Lagrangian $\mathcal{L}=\mathcal{L}(X, \phi)$ where $X= (1/2) \partial_\mu \phi \partial^{\mu}\phi$ as
    \begin{equation}
        c_e^2= \frac{\mathcal{L}_X}{\mathcal{L}_X+ 2 X\mathcal{L}_{XX}}
    \end{equation}
For canonical scalar fields $\mathcal{L}(X,\phi)=X-V(\phi)$, the effective speed of sound is one irrespective of the form of the potential \cite{R59a}. Since the canonical scalar field is considered in this $f(R,T)$ model, the effective sound speed will be one. Now, defining the gauge invariant scalar $\mathcal{R}$ called comoving curvature perturbation in $f(R,T)$ gravity as
\begin{equation}
    \mathcal{R}=\Phi- \frac{H}{\rho^{eff}+p^{eff}}\delta q
    \label{21k}
\end{equation}
where $\delta q$ is the scalar part of the 3-momentum density $(T^{eff})_i^0=\partial_i \delta q = -(1+2\lambda)\frac{\phi_0'}{a}\delta\phi$ and hence
\begin{equation}
    \mathcal{R}=\Phi
+ \frac{\mathcal{H}}{\phi'_0} \delta \phi
\label{21l}
\end{equation}
Upon substituting this in the Bardeen Eq. \ref{21h} and using Friedmann Eqs. \ref{e8} - \ref{e9}, Eq. \ref{21j} the evolution equation for the comoving curvature perturbation in Fourier space is derived as
\begin{equation}
     \mathcal{R}_k'= - \left( \frac{\mathcal{H}}{\mathcal{H}^2- \mathcal{H}'} \right) k^2 \Phi_k
     \label{21m}
\end{equation}
At super-Hubble scales the term $k^2\phi_k$ can be neglected and in that case $\mathcal{R}_k' \simeq 0$ which implies that the curvature perturbations are conserved when all the Fourier modes are outside the Hubble radius. Now, differentiating Eq. \ref{21m} w.r.t. conformal time and using the background Eqs. \ref{e8} - \ref{e9}, the Bardeen Eq. \ref{21h}, and Eq. \ref{21l}, the equation of motion governing the Fourier modes of the curvature perturbation induced by the scalar field is obtained as follows
\begin{equation}
    \mathcal{R}_k''+2\frac{z'}{z}\mathcal{R}_k' + k^2 \mathcal{R}_k = 0
    \label{21n}
\end{equation}
where $z$ is defined with the correction from the $f(R,T)$ gravity as
\begin{equation}
    z= \frac{a \phi_0'}{\mathcal{H}}\sqrt{1+2\lambda}
    \label{21o}
\end{equation}
 Then by introducing the Mukhanov-Sasaki variable in Fourier mode $v_k=z\mathcal{R}_k$, the Eq. \ref{21n} leads to the differential equation
\begin{equation}
    v_k''+ \left( k^2 - \frac{z''}{z} \right) v_k=0
    \label{21p}
\end{equation}
This is the Mukhanov-Sasaki equation in $f(R,T)$ gravity, which is identical to that in the GR case with the canonical scalar field. This equation can be rewritten in terms of the order of the Hankel function as 
\begin{equation}
    v_k''+ \left( k^2 - \frac{\nu_s^2-\frac{1}{4}}{\tau^2} \right) v_k=0
    \label{21q}
\end{equation}
where $\nu_s=\frac{3}{2}+3\epsilon_V -\eta_V$. The general solution to this equation is expressed as a linear combination of Hankel function as \cite{R5}
\begin{equation}
    v_k(\tau)=\sqrt{-\tau}\left[C_1H_{\nu_s}^{(1)} (-k\tau) +C_2H_{\nu_s}^{(2)}(-k\tau)  \right] \label{21r}
\end{equation}
Here $H_{\nu_s}^{(1,2)} (-k\tau)$ are the Hankel functions of the 1st and 2nd kind. $C_1$ and $C_2$ are constants to be evaluated using initial conditions.\\ \\
\textbf{At the sub-Hubble limit $(k \gg aH)$:} Initially, when all the quantum fluctuations are well inside the Hubble radius, it is assumed that the scalar perturbations are associated with Bunch-Davies vacuum and it requires $C_2=0$ \cite{R59e}. In this case $(x=-k\tau \rightarrow \infty)$,  the Hankel functions have the asymptotic limit \cite{R5}
\begin{equation}
    H_{\nu_s}^{(1,2)}(x) = \sqrt{\frac{2}{\pi x}} exp \left[\pm \left(x-\frac{\nu \pi}{2}-\frac{\pi}{4}\right)\right]
    \label{21s}
\end{equation}
This boundary condition implies that
\begin{equation}
    v_{k}(\tau)= \frac{\sqrt{-\pi\tau}}{2}e^{i(\nu_s+\frac{1}{2})\frac{\pi}{2}}H_{\nu_s}^{(1)} (-k\tau) 
    \label{21t}
\end{equation}
\textbf{At the supper-Hubble limit $(k \ll aH)$:} As the Universe expands during inflation, the physical wave length of the fluctuations grows larger than the Hubble horizon, and the quantum fluctuations become frozen as density perturbations. In this case $(-k\tau \rightarrow 0)$, the Hankel function reads \cite{R59e}
\begin{equation}
    H_{\nu_s}^{(1)} (-k\tau)= -\frac{i}{\pi}\Gamma(\nu)\left(\frac{-k\tau}{2}\right)^{-\nu}
    \label{21u}
\end{equation}
Then using Eq. \ref{21t} and Eq. \ref{21u}, the comoving curvature perturbation can be expressed as
\begin{equation}
    |\mathcal{R}_k|^2= \left[\frac{\Gamma(\nu)}{\Gamma(3/2)}(1-\epsilon_H)\right]^2 2^{2\nu -4} k^{-2\nu}(-\tau)^{3-2\nu}\frac{H^4}{(1+2\lambda)\dot\phi^2}
    \label{21v}
\end{equation}
When the curvature perturbations are quantized, the scalar power spectrum can be obtained as \cite{R59e, R59d}
\begin{align}
        \mathcal{P}_s(k) &= \left(\frac{k^3}{2\pi^2}\right) |\mathcal{R}_k|^2 \nonumber \\
        &= \left[\frac{\Gamma(\nu)}{\Gamma(3/2)} (1-\epsilon_H)\right]^2 \left(\frac{-k\tau}{2}\right)^{3-2\nu} \left(\frac{H^2}{2\pi \dot\phi \sqrt{1+2\lambda}}\right)^2
        \label{21w}
\end{align}
\textbf{At the Hubble horizon crossing $(k = aH)$:} The scalar power spectrum is then calculated just after the Hubble exit as 
\begin{equation}
    \mathcal{P}_s(k) = \left[\frac{\Gamma(\nu)}{\Gamma(3/2)} \right]^2 (1-\epsilon_H)^{2\nu-1} 2^{2\nu-3} \left(\frac{H^2}{2\pi \dot\phi \sqrt{1+2\lambda}}\right)^2
    \label{21z}
\end{equation}
At the leading order in the slow-roll approximation, the amplitude of the scalar spectra can be written from Eq. \ref{21z} as
\begin{align}
    \mathcal{P}_s(k) &= \frac{H^2}{4\pi^2}\frac{H^2}{(1+2\lambda)\dot\phi^2} \Bigg|_{k=aH} \nonumber \\
    &= \frac{H^2}{8\pi^2M_{Pl}^2 \epsilon_H} \Bigg|_{k=aH}
    \label{21a1}
\end{align}
The scalar spectral index is defined as \cite{R59d, R59e} 
\begin{equation}
    n_s= 1+ \frac{d \ln \mathcal{P}_s(k)}{d \ln k}
    \label{21a2}
\end{equation}
From Eq. \ref{21a1} and Eq. \ref{21a2}, the relation for the scalar spectral index in case of $f(R,T)$ gravity is obtained as 
\begin{equation}
    n_s=1- 6\epsilon_V+2\eta_V
    \label{21a3}
\end{equation}
Now in this $f(R,T)$ model, the gravitational action is linear in the Ricci scalar $R$ such that $f_R=\partial f / \partial R=1$ and hence the geometry part ($G_{\alpha\beta}$) of the modified Einstein field Eqs. \ref{e5} remains unaltered. Further, the correction term in the action represents a minimal coupling to the trace of the EM tensor, and it does not contribute to the anisotropic stress at the linear level. Since the tensor perturbations arise completely from the $g_{\alpha\beta}$ fluctuations in the curvature sector of the action, the evolution equations for the tensor modes $h_{ij}$ in the $f(R,T)$ gravity will be identical to that of GR. Therefore, the amplitude of the tensor spectra reads \cite{R59d}
\begin{equation}
    \mathcal{P}_t(k)= \frac{8}{M_{Pl}^2}\frac{H^2}{4\pi^2} \Bigg|_{k=aH}
    \label{21a4}
\end{equation}
Therefore, the tensor-to-scalar ratio in slow roll limit in $f(R,T)$ gravity is found to be
\begin{equation}
    r= \frac{\mathcal{P}_t(k)}{\mathcal{P}_s(k)} = 16 \epsilon_V
    \label{21a5}
\end{equation}
The running of the spectral tilt is defined as \cite{R59e}
\begin{equation}
    n_{sk}=\frac{d n_s}{d \ln k} \Bigg|_{k=aH} 
    \label{21a6}
\end{equation}
In terms of slow roll parameter, the running of the scalar spectral tilt in $f(R,T)$ gravity can be expressed as
\begin{equation}
    n_{sk} = 16\epsilon_{V}\eta_{V}-24\epsilon_{V}^2 -2\xi_{V}^2
    \label{21a7}
\end{equation}
It is clear that the expressions for the scalar spectral tilt (Eq. \ref{21a3}), the tensor-to-scalar ratio (Eq. \ref{21a5}), and the running of the scalar spectral tilt (Eq. \ref{21a7}) remain unaltered in case of $f(R,T)$ gravity. However, the modifications due to the trace term are encapsulated within the definition of the potential slow-roll parameters.

\section{Models under study} \label{s4}
\subsection{Mutated hilltop Inflation (MHI)}
The mutated hilltop potential is a refined version of the standard hilltop inflation potential, developed by Pal et al. \cite{R60} In case of the mutated version, the flat potential is altered by a hyperbolic function whose power series expansion includes an infinite number of terms which makes the model more accurate and concrete compared to the traditional hilltop potential. The form of this potential bears close similarity with its counterpart in mutated hybrid inflation, hence the name \cite{R60,R61,R62}. The form of the potential reads
\begin{equation}
    V(\phi)=V_{0} [ 1- \sech{(\chi \phi)} ]
    \label{e22}
\end{equation}
where $V_0$ is the inflation energy scale and $\chi$ is a parameter with the dimension of inverse of $Planck$ mass. The $\chi$ flattens the potential near hilltop, thereby creating a longer period for sufficient $e$-folds can happen.  \\
This potential is purely phenomenological, however, motivated from supergravity \cite{R60,R63}. MHI is distinct from HI in two ways which are as follows:
\begin{itemize}
    \item In hilltop potential, inflation takes place near the maxima of the potential $(\phi \approx 0)$ where as in the case of mutated hilltop potential, inflation happens on a flat plateau away from the minimum, i.e. as the inflaton rolls down towards it's minima. 
    \item MHI satisfies the condition of vanishing of the potential  at its absolute minimum, i.e. $V (\phi_{min}) = V^\prime (\phi_{min}) = 0$, which is not seen in the case of usual hilltop potential.
\end{itemize}
Mutated hilltop potential generally produces small tensor-to-scalar ratio in GR \cite{R60,R62}. Mayukh et al. reported that in Einstein-Gauss-Bonnet gravity, it produces small $r$ compatible with current observation \cite{R64}. Keeping this point in mind, we will check whether this potential is viable in the $f(R,T)$ gravity framework. \\
The slow roll parameters for this potential can be expressed using Eqs. \ref{e17}, \ref{e18} as
\begin{equation}
    \epsilon_V = \frac{M_{Pl}^2 \chi^2 \sech^2{(\chi \phi)} \tanh^2(\chi \phi)} {2 (1+2\lambda)[1-\sech{(\chi \phi)}]^2}
    \label{e23}
\end{equation}
\begin{equation}
    \eta_V= \frac{M_{Pl}^2 \chi^2[\sech^3{(\chi \phi)} -  \sech{(\chi \phi)} \tanh^2{\chi\phi}]}{(1+2\lambda)[1-\sech{(\chi \phi)}]}
    \label{e24}
\end{equation}
\begin{equation}
    \xi_{V}^2=\frac{M_{Pl}^4\chi^4 \coth^2{(\frac{\chi\phi}{2})} \sech^4{(\chi\phi)} [\cosh{(2\chi\phi)}-1]}{2(1+2\lambda)^2} 
    \label{e25}
\end{equation}
Then the scalar spectral index and tensor-to-scalar ratio can be calculated from Eq. \ref{21a3}, \ref{21a5} as
\begin{equation}
    n_s= 1- \frac{3 M_{Pl}^2 \chi^2 \sech^2{(\chi \phi)} \tanh^2(\chi \phi)} { (1+2\lambda)[1-\sech{(\chi \phi)}]^2} + \frac{2 M_{Pl}^2 \chi^2[\sech^3{(\chi \phi)} -  \sech{(\chi \phi)} \tanh^2{\chi\phi}]}{(1+2\lambda)[1-\sech{(\chi \phi)}]}
    \label{e26}
\end{equation}
\begin{equation}
    r= \frac{8M_{Pl}^2 \chi^2 \sech^2{(\chi \phi)} \tanh^2(\chi \phi)} { (1+2\lambda)[1-\sech{(\chi \phi)}]^2}
    \label{e27}
\end{equation}
The expression for $n_{sk}$ is skipped, being too lengthy, however can be easily calculated using Eqs. \ref{21a7}, \ref{e23}, \ref{e24} and \ref{e25}.  The $n_s-r$ plot for the mutated hilltop potential is presented below for different values of $\lambda$ at 60 $e$-folds and keeping $\chi$ at $3M_{Pl}^{-1}$ scale.
\begin{figure}
    \centering
    \includegraphics[width=0.5\linewidth]{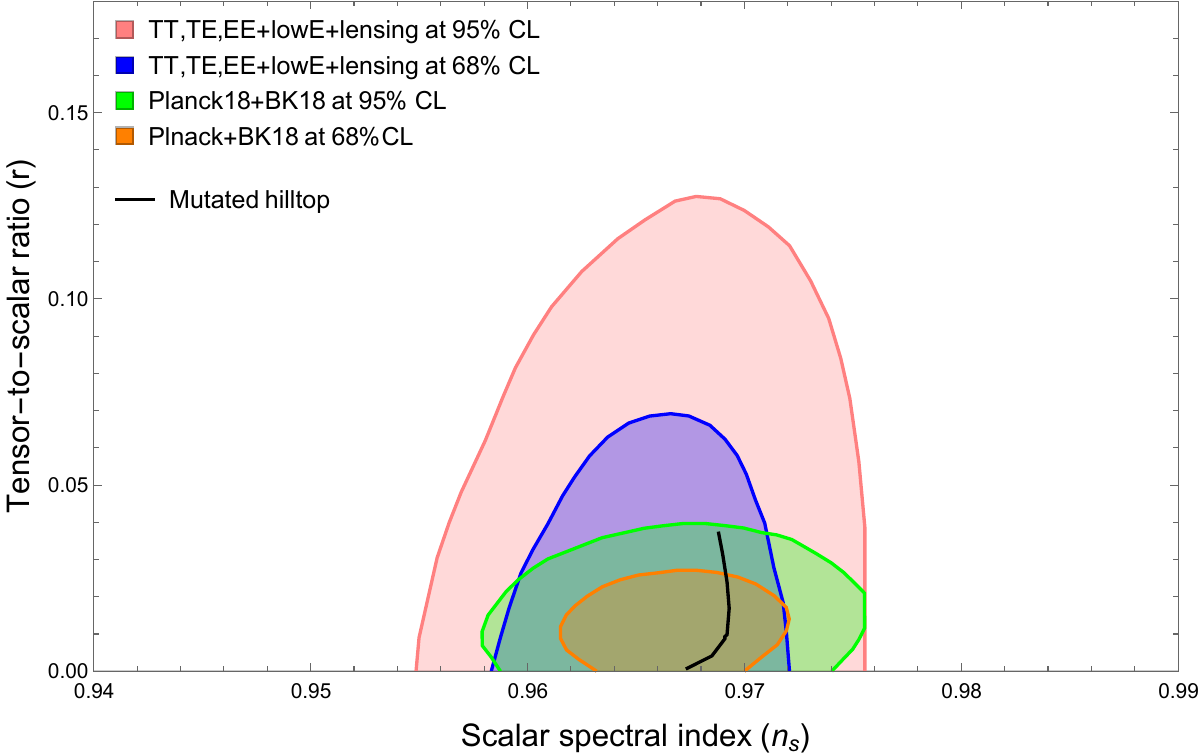}
    \caption{The $n_s-r$ plot predicted by the mutated hilltop potential. The marginalized joint 68\% and 95\% C.L. regions for $n_s$ and $r$ at $k=0.002 Mpc^{-1}$ from $Planck$ 2018 data are shown in blue and pink colour where as for $Planck$+BK18 are shown in orange and green respectively.}
    \label{f1}
\end{figure}
It is clear from Figure \ref{f1} that the trajectory in the $n_s-r$ plane enters the $1\sigma$ region of $Planck$ 2018 and almost covers the entire $1\sigma$ region of $Planck$+BK18. It is found that with a suitable choice of the model parameter $\lambda$, the tensor-to-scalar ratio can be achieved of the order $10^{-4}$ that not only matches the current bounds, but will also be aligned with the future expected bounds of LiteBIRD \cite{R65}, CMB-S4 \cite{R66}. For example, when $\lambda=0.5$, we have $n_s=0.9671$ and $r=4.7 \times 10^{-4}$ respectively. In addition, the range of model parameters is determined to be $0.1 < \lambda < 29$ considering $r < 0.036$. Then running $n_{sk}$ is found to be negative and very small up to the order of $10^{-4}$, the numerical value is listed in Table \ref{T1}.


\subsection{D-Brane Inflation}
\begin{figure}[h!]
 \centering
 \begin{subfigure}[h]{0.48\textwidth}
     \centering
     \includegraphics[width=\textwidth]{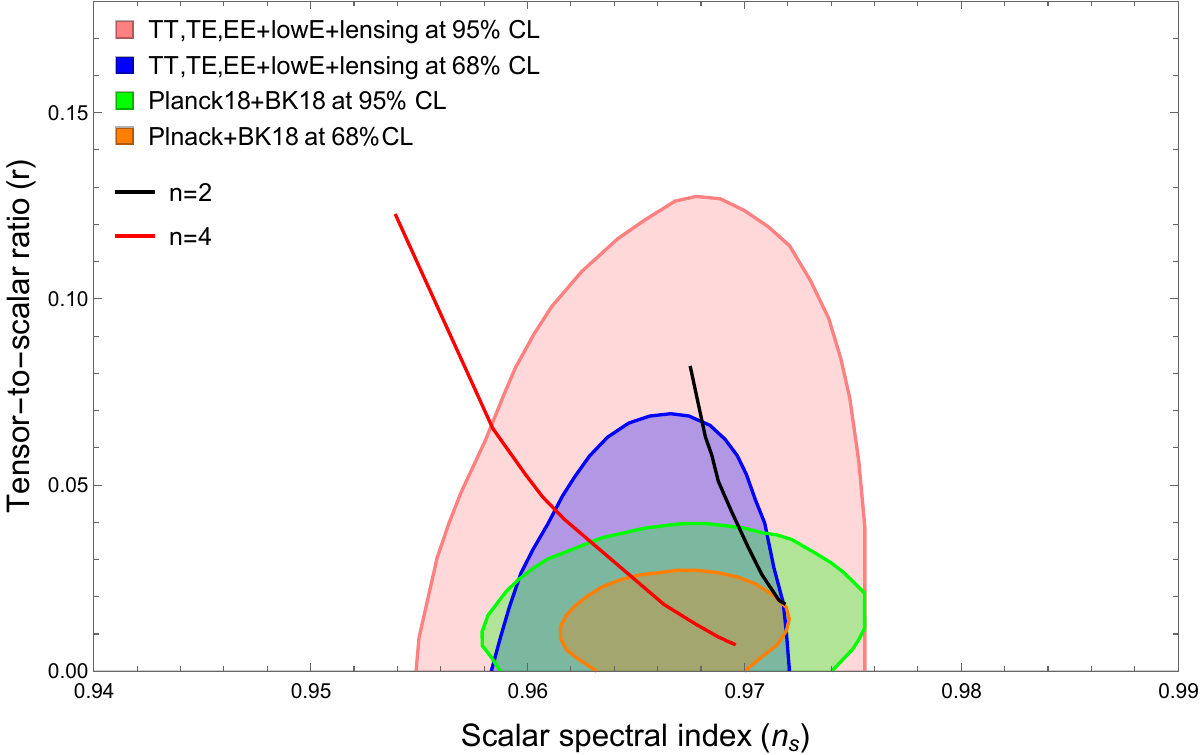}
     \caption{for $m= 5$}
     \label{2a}
 \end{subfigure}
 \hfill
 \begin{subfigure}[h]{0.48\textwidth}
     \centering
     \includegraphics[width=\textwidth]{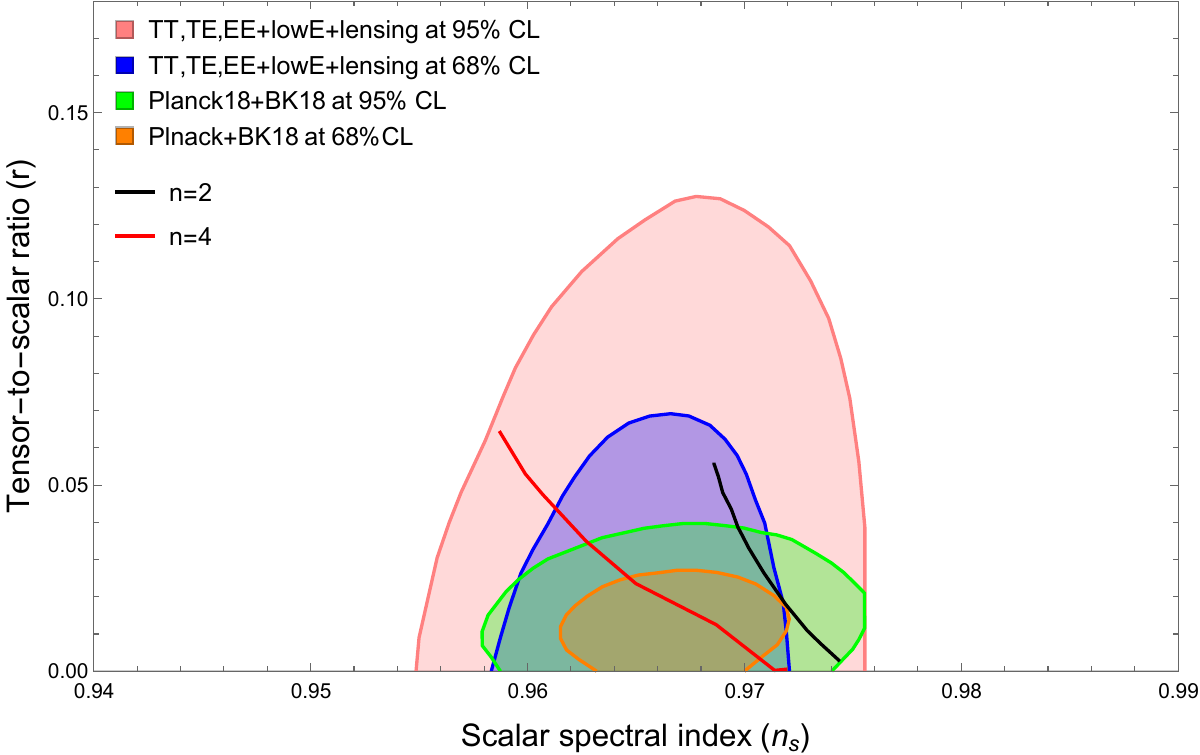}
     \caption{for $m=0. 5$}
 \end{subfigure}
 \caption{ \small The $n_s-r$ trajectory predicted by the KKLT potential at 60 $e$-folding. The marginalized joint 68\% and 95\% C.L. regions for $n_s$ and $r$ at $k=0.002 Mpc^{-1}$ from $Planck$ 2018 data are shown in blue and pink colour where as for $Planck$+BK18 are shown in orange and green respectively. }
    \label{f2}
 \end{figure}

The concept of brane inflation was proposed by Dvali et al. in 1999 where they considered the separation between a brane and an anti-brane as the inflaton \cite{R67}. As the branes move close to each other, slow-roll inflation occurs, and when the branes collide, energy is released into radiation, which corresponds to the reheating mechanism \cite{R67}. \\
D-Brane inflation is a specific type of brane inflation, involving D-branes that obey the Dirichlet boundary condition. D-brane inflation involves $D3$ and Anti-$D3$ brane placed in a wrapped geometry. The potential energy from the tensions of the $D3/\overline{D3}$ branes drives the inflation. The brane inflation scenario requires moduli stabilization so that they can realize the observable universe in 4-dimensions. The Kachru-Kallos-Linde-Maldacena-McAllister-Trivedi (KKLMMT) mechanism of adding an Anti-$D3$ brane to uplift the vacuum to a metastable de-Sitter is an example of D-brane inflation with moduli stabilization \cite{R68}. \\
Despite theoretical issues, D-brane inflation gained limelight after the release of $Planck$ 2018, as it matches the current bounds \cite{R9}. We will focus on two types of D-brane model using the nomenclature used by Jerome et al. \cite{R8}. The first model is BI, which stands for brane inflation, has the Coulomb-type potential:
\begin{equation}
    V_{BI}=V_0 \left(1- \frac{m^n}{\phi^n} \right)
    \label{e28}
\end{equation}
Although this potential matches the current bound appropriately, it has certain issues. This potential is unbounded from below and hence stucked with graceful exit problem \cite{R69}. KKLMMT proposed a consistent generalization in which the potential takes the form of an inverse harmonic function \cite{R68}. This is named as KKLTI by Jerome et al. \cite{R8} which stands for KKLT inflation and has the potential: 
\begin{equation}
    V_{KKLTI}=V_0 \left(1- \frac{m^n}{\phi^n} \right)^{-1} = V_0 \frac{\phi^n}{\phi^n + m^n}
    \label{e29}
\end{equation}
Since $V_{BI}$ is not physically sound, we will continue our further study with $V_{KKLTI}$ model. In the large $m$ limit ($m\gg10$), $V_{KKLTI}$ predicts a large $r=8n/N$, which is not supported by current observation. However, in the small $m$ limit ($m\le1$), it provides a good fit for $Planck$ 2018. The case $n=2$ covers the entire $1\sigma$ region of $Planck$ \cite{R69}. Now, keeping this in mind, we aim to check how the KKLTI model behaves in the $f(R,T)$ gravity setup both in the large and small $m$ limits. \\
The potential slow-roll parameters are obtained as
\begin{equation}
    \epsilon_V=\frac{M_{Pl}^2 n^2 m^{2n}}{2(1+2\lambda)\phi^2(\phi^n+m^n)^2}
    \label{e30}
\end{equation}
\begin{equation}
    \eta_V=\frac{M_{Pl}^2nm^n[(n-1)m^n-(n+1)\phi^n]}{(1+2\lambda)\phi^2(\phi^n+m^n)^2}
    \label{31}
\end{equation}
\begin{equation}
    \xi_{V}^2= \frac{M_{Pl}^4 n^2m^{2n}[m^{2n}(n^2-3n+2)-4(n^2-1)m^n\phi^{n}+(n^2+3n+2)\phi^{2n}]}{(1+2\lambda)^2\phi^4(\phi^n+m^n)^4}
    \label{e32}
\end{equation}
The scalar spectral index, tensor-to-scalar ratio, and running of ths scalar spectral index are expressed as:
\begin{equation}
    n_s= 1-\frac{M_{Pl}^2 n m^n [(n+2)m^n+2(n+1)\phi^n]}{(1+2\lambda)\phi^n(\phi^n+m^n)^2}
    \label{e33}
\end{equation}
\begin{equation}
    r= \frac{8M_{Pl}^2 n^2 m^{2n}}{(1+2\lambda)\phi^2(\phi^n+m^n)^2}
    \label{e34}
\end{equation}
\begin{equation}
    n_{sk}=-\frac{2M_{Pl}^4 n^2m^{2n}[(n+2)m^{2n}+4(n+1)m^n\phi^n+(n^2+3n+2)\phi^{2n}]}{(1+2\lambda)^2\phi^4(\phi^n+m^n)^4}
    \label{e35}
\end{equation}
The $n_s-r$ trajectory for the $V_{KKLTI}$ potential is presented for different values of $\lambda$ in Figure \ref{f2}. We restricted our study to two particular cases $n=2$ \& $n=4$ for $m=0.5$ and $m=5$. Let us now discuss the results found in the study.
\begin{itemize}
    \item At large $m=5$ limit, both the trajectories for $n=2$ and $n=4$ enter the $1\sigma$ limit of $Planck$ 2018 for certain values of the model parameter. While $n=4$ enters the $1\sigma$ region of $Planck$+BK18, $n=2$ only touches the boundary. The model parameter space is obtained to be $2.3<\lambda<3.7$ and $0.11<\lambda<3.6$ for $n=2$ and $n=4$, respectively at $N=60$ $e$-folding considering $r< 0.056$. The running $n_{sk}$ is found to be negative on the order of $10^{-4}$, perfectly aligned with $Planck$.
    \item At small $m=0.5$ limit, a similar trend is observed for both cases. $n=4$ line covers entire $1\sigma$ region of both $Planck$ 2018 and $Planck$+BK18. While $n=2$ enters $1\sigma$ of $Planck$ and $2\sigma$ of $Planck$+BK18 thereby touching the $1\sigma$ boundary. For the bound $r< 0.056$, the viable parameter space is found to be $285<\lambda<414$ and $60<\lambda<406$ for $n=2$ and $n=4$ respectively, at $N=60$ $e$-fold. A small negative running $(n_{sk} \sim 10^{-4})$ is obtained.
    
\end{itemize}

\subsection{Woods-Saxon Inflation}

\begin{figure}[h!]
 \begin{subfigure}[h]{0.48\textwidth}
     \centering
     \includegraphics[width=\textwidth]{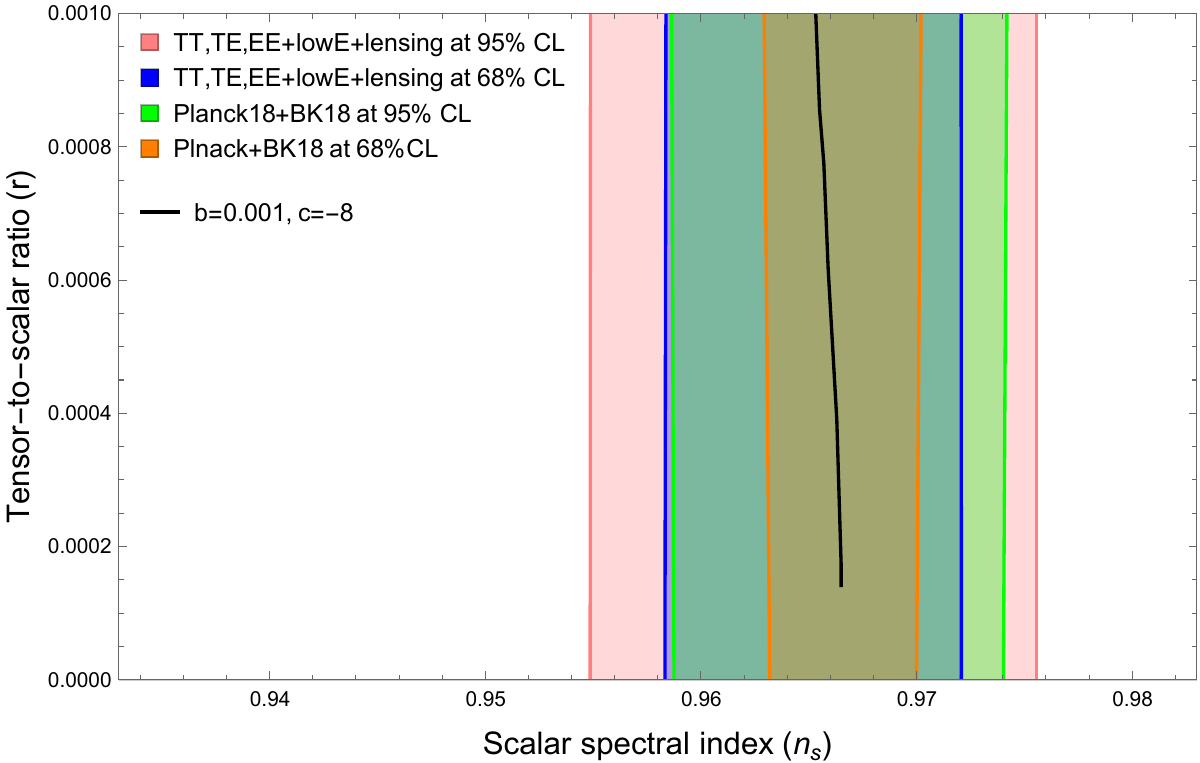}
     \label{3a}
 \end{subfigure}
 \hfill
 \begin{subfigure}[h]{0.48\textwidth}
     \centering
     \includegraphics[width=\textwidth]{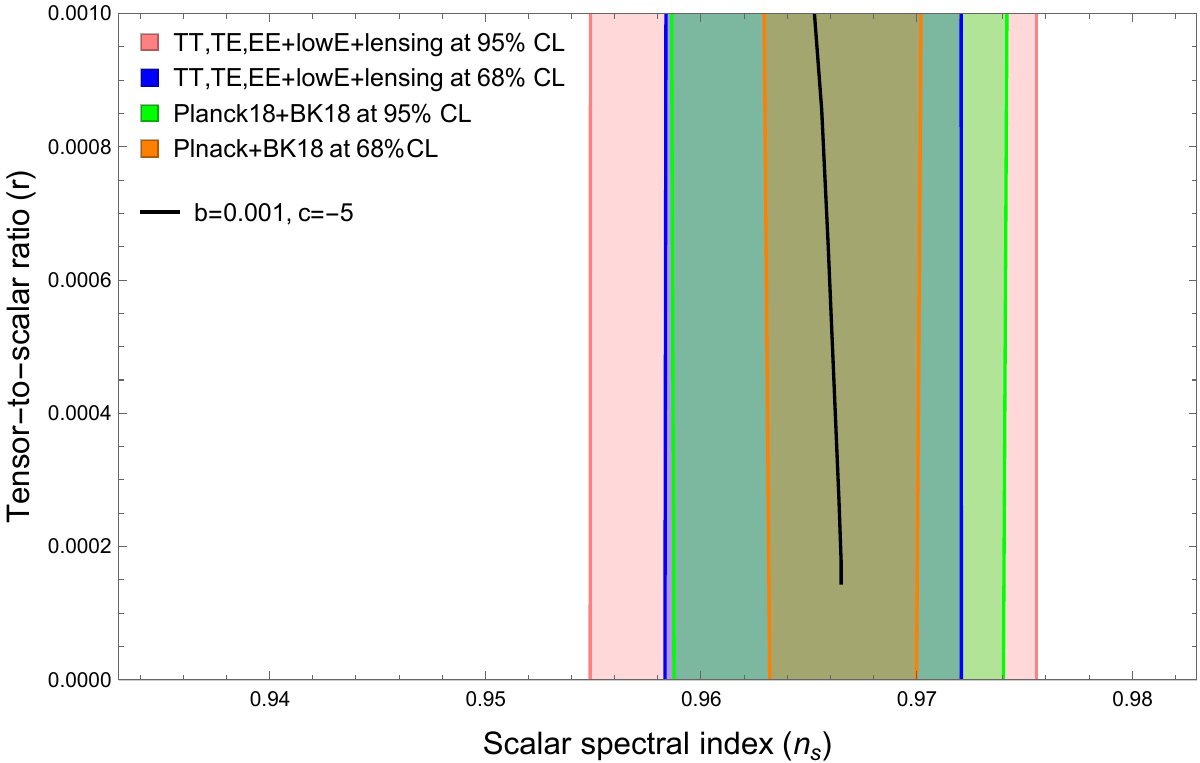}
     \label{3b}
 \end{subfigure}
 \caption{ \small The $n_s-r$ trajectory predicted by the Woods-Saxon potential at 60 $e$-folding. The marginalized joint 68\% and 95\% C.L. regions for $n_s$ and $r$ at $k=0.002 Mpc^{-1}$ from $Planck$ 2018 data are shown in blue and pink colour where as for $Planck$+BK18 are shown in orange and green respectively. }
    \label{f3}
 \end{figure}

Wood-Saxon inflation is based on a potential that was originally used to describe the nuclear force experienced by the nucleons in the shell model \cite{R70}. However, it has been used in cosmological inflation by Oikonomou et al., since the potential has a flat plateau ideal for slow roll inflation \cite{R71}. \\ 
Woods-Saxon potential reads as
\begin{equation}
    V(\phi)=\frac{V_0}{1+be^{-c\kappa\phi}}
    \label{e36}
\end{equation}
where $V_0$ is the depth of the potential with dimension $eV^4$. The free parameter $c$ has the dimension of $eV$, and $b$ is dimensionless. Whereas $\kappa$ is a constant. \\
In GR, Oikonomou et al. showed that the potential meets the $Planck$ bound for $b > 0$ and the varying $a < 0$. In addition, the reheating era can be successfully described with the same values of the free parameters. However, the parameter $V_0$ affects only the reheating era \cite{R71}. Venikoudis et al. showed that in Gauss-Bonnet gravity, the Woods-Saxon potential successfully explain both the inflation and reheating eras \cite{R72}. A similar result was reported by Younesizadeh et al. with a different Gauss-Bonnet coupling function \cite{R73}. In addition, Radhakrishnan et al. reported that the Woods-Saxon model can be considered as a potential alternative to $\Lambda$CDM to explain late-time acceleration \cite{R74}. These results motivate us to explore the theoretical predictions of this potential in $f(R,T)$ gravity. \\
Then the potential slow-roll parameters in $f(R,T)$ gravity read as
\begin{equation}
    \epsilon_{V}=\frac{b^2c^2e^{-2c\kappa\phi}}{2(1+2\lambda)(1+be^{-c\kappa\phi})^2}
    \label{e37}
\end{equation}
\begin{equation}
    \eta_V=\frac{bc^2(b-e^{c\kappa\phi})}{(1+2\lambda)(b+e^{c\kappa\phi})^2}
    \label{e38}
\end{equation}
\begin{equation}
    \xi_{V}^2=\frac{b^2c^4(b^2-4be^{c\kappa\phi}+e^{2c\kappa\phi})}{(1+2\lambda)^2(b+e^{c\kappa\phi})^4}
    \label{e39}
\end{equation}
Then the expressions for scalar spectral index, tensor-to-scalar ratio, and running read as
\begin{equation}
    n_s=\frac{2be^{c\kappa\phi}(1-c^2+2\lambda)+e^{2c\kappa\phi}(1+2\lambda)+b^2(1-c^2+2\lambda)}{(1+2\lambda)(b+e^{c\kappa\phi})^2}
    \label{e40}
\end{equation}
\begin{equation}
    r= \frac{8b^2c^2 e^{-2c\kappa\phi}}{(1+2\lambda)(1+be^{-c\kappa\phi})^2}
    \label{e41}
\end{equation}
\begin{equation}
    n_{sk}=-\frac{2b^2c^4 e^{2c\kappa\phi}}{(1+2\lambda)^2(b+e^{c\kappa\phi})^4}
    \label{e42}
\end{equation}
The trajectory of the potential in the $n_s-r$ plane is presented in Figure \ref{f3}. To do so, $\kappa$ is set to unity. The free parameters $b, c$ are very sensitive and are fixed at $b=0.001$ and $c=-5,-8$, respectively. Considering $N=60 $ $e$-folds, the model parameter $\lambda$ is varied to see how it affects the results.
\begin{itemize}
    \item The line for $c=-5$ enters the $1\sigma$ region of the $Planck$+BK18 bound. The model can achieve $r$ of the order of $10^{-4}$ for different values of $\lambda$. The parameter space of the model is found to be $0.2< \lambda < 5.8$. Then running $n_{sk}$ is also found to agree with $Planck$. 
    \item A similar trend is seen for the case $c=-8$. The trajectory enters the $1\sigma$ bound of $Planck$+BK18  for a small window of $1.4 < \lambda < 15.6$. Very small tensor-to-scalar ratio and negative running are achieved. The summarized results are shown in Table \ref{T1}.
\end{itemize}

\begin{table}[htbp]
    \centering
    \begin{tabular}{lccccl}
		\toprule
		Inflationary Model & Fixed Parameter & \(\lambda\)  & $n_s$ & $r$ & \hspace{0.7cm} $n_{sk}$ \\
		\midrule
		
		KKLT ($n=2$) & $m=0.5$  & 310 &  0.9690 & 0.048 & $-5.36 \times10^{-4}$\\
        
		KKLT ($n=2$) & $m=5$  & 3 & 0.9688 & 0.051  &  $-5.39 \times10^{-4}$\\
        
        KKLT ($n=4$) & $m=0.5$ & 200 & 0.9650 & 0.023  & $-6.57 \times 10^{-4}$\\
        
		KKLT ($n=4$) & $m=5$ & 0.5 &  0.9678 & 0.012  & $-5.80 \times10^{-4}$
         \\
Mutated hilltop & $\chi=3$  & 1 & 0.9673 & $6.99 \times10^{-4}$   &   $-5.36 \times 10^{-4}$\\
         
        Woods-Saxon & $b=0.001$, $c=-5$ & 3 &  0.9659 & $6.53 \times 10^{-4}$ & $-5.57 \times 10^{-4}$ \\
        Woods-Saxon & $b=0.001$, $c=-8$ & 10 & 0.9657 &  $7.75 \times 10^{-4}$ & $-5.83 \times10^{-4}$ \\
		\bottomrule
	\end{tabular}

    \caption{Theoretical predictions by the models under study for different observables calculated at $N=60$ $e$-folds }
    \label{T1}
\end{table}


\subsection{Models in light of P-ACT-LB-BK18} \label{s4c}
\begin{figure}[h!]
    \centering
    \includegraphics[width=0.5\linewidth]{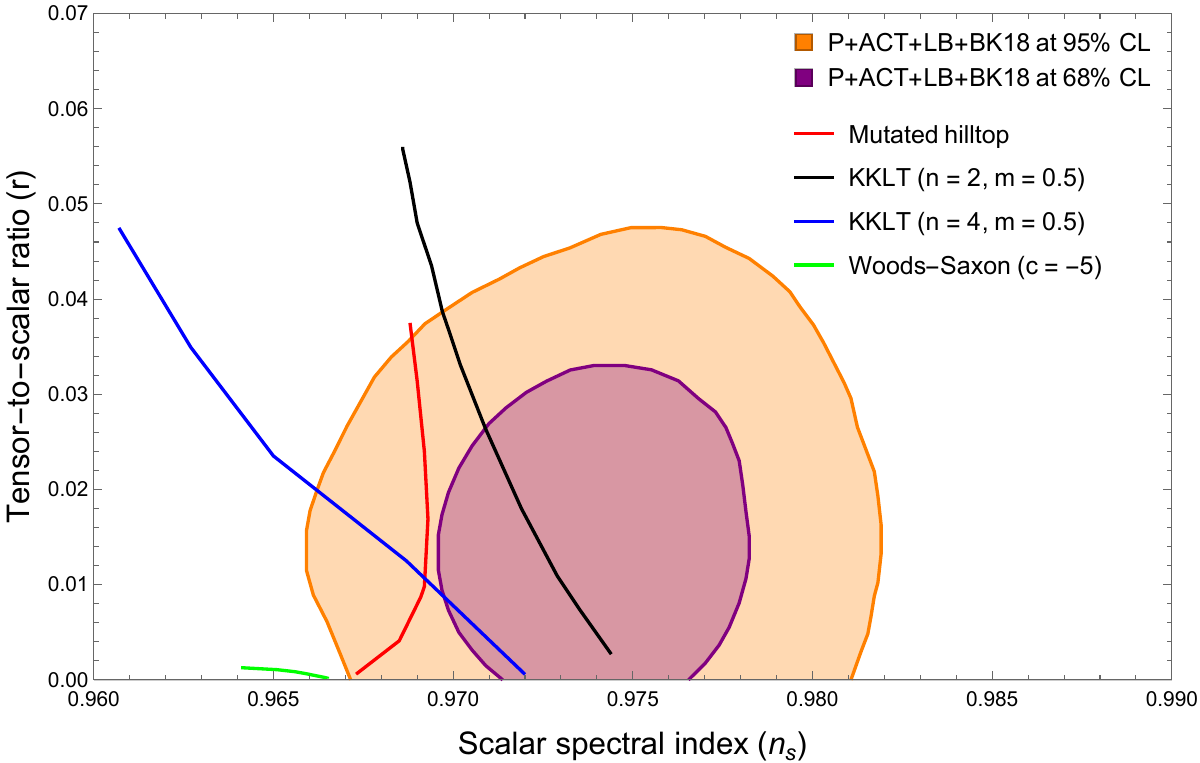}
    \caption{\small The $n_s-r$ trajectory predicted by the mutated hilltop potential, KKLT $(n=2, m=0.5)$, KKLT $(n=4, m=0.5)$, Woods-Saxon potential $(b=0.001, c=-5)$ at 60 $e$-folding are shown in red, black, blue, and green colour respectively. The marginalized joint 68\% and 95\% C.L. regions for $n_s$ and $r$ at $k=0.002 Mpc^{-1}$ from the combined P-ACT-LB-BK18 data are shown in purple and orange colour. }
    \label{fig:placeholder}
\end{figure}

The recent data release by the Atacama Cosmology Telescope developed a new tension on the value of the scalar spectral tilt. The new value reported by ACT DR6 when combined with the BAO data of DESI DR2 stands $2\sigma$ away from $Planck$ value in the $n_s-r$ plane \cite{R11,R12}. In addition, the Atacama dataset in combination with BICEP/$K$eck provides larger upper limit of the tensor-to-scalar ratio \cite{R12}. The reported values read
\begin{equation*}
    n_s = 0.9743\pm 0.0034 \hspace{1cm} \text{(P-ACT-LB)}
   \end{equation*}
\begin{equation*}
     r < 0.0038 \hspace{1cm} \text{(95\%, P-ACT-LB-BK18)}
\end{equation*}
These elevated ranges of $n_s$ and $r$ will have a severe effect on the viability of any inflationary model. The three models studied in the previous sections were found to be consistent with the CMB bounds of $Planck$ and BICEP/$K$eck in $f(R,T)$ gravity. Now, we will check whether they can meet the bound provided by the P-ACT-LB-BK18. The trajectories of the three potentials in the $(n_s-r)$ plane are presented in Figure \ref{fig:placeholder}. \\ \\
The mutated hilltop potential stays within the $2\sigma$ region of the combined P-ACT-LB-BK18 bound. Both KKLT trajectories for cases $(n=2, m=0.5)$ and $(n=4, m=0.5)$ successfully cover the $1\sigma$ contour. However, the Woods-Saxon potential is found to remain outside the Atacama contours. In short, the mutated hilltop and KKLT model are viable to the recent P-ACT-BL-BK18 bound in $f(R,T)$ gravity.

\section{Conclusion} \label{s5}
Its been five decades, the theoretical framework for inflation is continuously developing to describe the primordial universe, both in general relativity and in modified gravity theories. From scalar fields to higher-dimensional string theory, from super gravity to non-minimal coupling of matter with fields, inflationary dynamics has been explored to meet the observational results. In this study, we performed a detailed numerical analysis of three well-motivated inflationary models, viz. mutated hilltop inflation, D-brane inflation, and Woods-Saxon inflation within $f(R,T)$ gravity. The simplest yet most used form $f(R,T)=R+16\pi G \lambda T$ is undertaken, and the slow-roll inflation is studied. The scalar and tensor power spectra are derived to calculate the scalar spectral tilt, the tensor-to-scalar ratio, and the running of the scalar spectral index. The recent bounds of $Planck$, BICEP/$Keck$, ACT DR6, and DESI DR2 have been used to confront the theoretical prediction of these three models. In addition, it is also checked whether the models can predict the projected sensitivities of the upcoming CMB probes like LiteBIRD, CMB-S4 etc.  \\ \\
\textbf{Mutated hilltop Inflation} is found to retain its excellent observational prediction up to $r \sim 10^{-4}$ in $f(R,T)$ gravity for a suitable choice of the model parameter $\lambda$. Its prediction remains firmly within the CMB $1 \sigma$ contour of $Planck$+BK18, and in the $2\sigma$ region of P-ACT-LB-BK18. The parameter space is calculated as $0.1 < \lambda < 29$ and the running is well within the $Planck$+BK15 range with a small negative value. \\ \\
\textbf{D-brane Inflation} arising from the $D3/\overline{D3}$ interaction is found to be consistent with the current data. The KKLTI model is considered with specific cases $n=2$ and $n=4$. For both large $m=5$ and small $m=0.5$ limits, these models exhibit moderate tensor signatures $(r \sim 10^{-2} $ to $10^{-3})$ and predict suppressed negative $n_{sk} \sim 10^{-4}$ in $f(R,T)$ gravity. The $n=4$ trajectory stays in the $1 \sigma$ contour of $Planck$+BK18 for the large and small $m$ limits, while the $n=2$ model touches the $1 \sigma$ boundary. Both models are consistent with the combined bound of $Planck$+BK18, ACT-DR6 and DESI-DR2 at the $1\sigma$ confidence level. The parameter space evaluated for each case considering $r < 0.056$  is summarized in Table \ref{T2}. \\ \\
\textbf{Woods-Saxon Inflation}, motivated from nuclear shell model, predicts a small tensor-to-scalar ratio $(r \sim 10^{-4})$ well below the current detection threshold. The scalar spectral index spans $n_s \in [0.964, 0.966]$ as $\lambda$ varies and the running $n_{sk}$ is negative on the order of $10^{-4}$. The trajectory remains inside the $1\sigma$ contour of $Planck$+BK18, while it stays outside the P-ACT-LB-BK18 contour. Finally, the model parameter space is evaluated as $0.2 < \lambda < 5.8$ when $c=-5$ and $1.4 < \lambda < 15.6$ when $c=-8$. \\ 

\begin{table}[htbp]
	\centering
	\begin{tabular}{lcccl}
		\toprule
		Model & Fixed parameters  & Allowed range & Remark ($r$) & Compatible with\\
		 \midrule
        Mutated hilltop & $\chi=3$  & $0.1<\lambda<29$ & $r \sim 10^{-4}$  & $Planck$, BICEP/$Keck$, \\
        &&&& ACT DR6, DESI DR2 \\  \midrule
	  KKLT ($n=2$) & $m=5$ & $2.3<\lambda<3.7$ &  & \\
	  KKLT ($n=2$) & $m=0.5$  & $285<\lambda<414$ &  $r \sim 10^{-2}$ & $Planck$, BICEP/$Keck$, \\
	  KKLT ($n=4$) & $m=5$  & $0.11<\lambda<3.6$ &  to $10^{-3}$ & ACT DR6, DESI DR2 \\
	  KKLT ($n=4$) & $m=0.5$ & $60<\lambda<406$ &   & 
         \\  \midrule
        Woods-Saxon & $b=0.001$, $c=-8$ & $1.4<\lambda<15.6$ &  $r \sim 10^{-4}$  & $Planck$, BICEP/$Keck$\\
         & $b=0.001$, $c=-5$ & $0.2<\lambda<5.8$ & $r \sim 10^{-4}$ \\
		\bottomrule
	\end{tabular}
    \caption{Summary of parameter space evaluated for models under study at $n=60$ $e$-folds with fixed parameters and their compatibility with different CMB probes }
	\label{T2}
\end{table}
Another important issue is to be mentioned that the vacuum expectation value $(vev)$ of the inflaton requires to be sub-$Planckian$ to avoid theoretical clashes with particle physics. In the $f(R,T)$ gravity framework, it is seen that the inflationary models under study remain consistent with the current CMB bounds at and below the $Planck$ scale. Again, both the mutated hilltop and the Woods-Saxon potential produce an extremely small tensor-to-scalar ratio $(r \sim 10^{-4})$ in $f(R,T)$ gravity; hence, they will be aligned with future CMB probes such as LiteBIRD, CMB-S4. As a possible extension of this work, one can study different other inflationary models within $f(R,T)$ gravity considering both the minimal and non-minimal coupling between matter and metric. In addition, Palatini and Metric-affine versions of ${f(R,T)}$ gravity may be checked to study inflationary models, which may generate interesting results.

\section*{Acknowledgments}
The authors would like to thank Prof. David Wands, University of Portsmouth for insightful conceptual discussion and valuable suggestions regarding perturbation calculation presented in the manuscript.

\printbibliography

\end{document}